\documentclass[11pt]{revtex4}   
\usepackage{graphicx}
\usepackage[ansinew]{inputenc} 
\begin{document}

\bibliographystyle{apsrev}

\title{Nonlocal reflection by photonic barriers}
\author{R.--M.~Vetter, A.~Haibel, and G.~Nimtz}
\affiliation{Universität zu Köln, II.~Physikalisches Institut, Zülpicher Str.\,77, 
                      D-50937 Köln, Germany}

\begin{abstract}  
The time behaviour of microwaves undergoing partial reflection by photonic barriers was measured 
in the time and in the frequency domain. It was observed that unlike 
the duration of  partial reflection by dielectric layers, the measured reflection duration 
of barriers is independent of their length. The experimental results point to a nonlocal 
behaviour of evanescent modes at least over a distance of some 
ten wavelengths. Evanescent modes correspond to photonic tunnelling in quantum mechanics.
\end{abstract}

\maketitle

\section{Introduction} 

We are used to measuring a reflection time from partial 
reflection of light, for instance by a sheet of glass, that is determined by sheer thickness.
 The reflection is observed 
only after a time span corresponding to twice the layer thickness multiplied by the group 
velocity of light in glass. 
Three hundred years ago Newton conjectured that light was composed of corpuscles and 
argued rather in the case of partial reflection by two or more surfaces: "Light striking the first surface 
sets off a kind of wave or field that travels along with the light and predisposes it to reflect or 
not reflect off  the second surface." He called this process 'fits of easy reflection or easy 
transmission'~\cite{Feynman}. As theory and experiments have shown this is not correct.
In the case of dielectric media with a positive real part of the refractive index~$n$ like glass, the 
reflection is composed of components from both the front and the back surface reflection of the sheet of glass. 

Amazingly, in the case of reflection of particles by an opaque photonic barrier, where the   
index of refraction is purely imaginary, Newton's conjecture seems to be close to reality: Partial 
reflection by opaque photonic barriers suffers a short but constant time delay independent of the 
barrier's length. A barrier is called opaque if its transmission is less than $1/e$. For the photonic barriers investigated here, the incident particles can be 
simulated by localized wave packets. We found that for these wave packets the reflection time 
equals the transmission time observed in photonic tunnelling experiments~\cite{Haibel}.
This behaviour is 
opposite to the partial reflection of dielectric sheets and may be explained by 
a nonlocality of evanescent modes in opaque barriers. 
Nonlocality and causality 
were investigated in Ref.~\cite{hegerfeldt} and quite recently with respect to superluminal photonic 
tunnelling by Perel'man \cite{perel}.

\section{Experimental setup}

\begin{figure}[hbt]
 \includegraphics[width=0.6\linewidth]{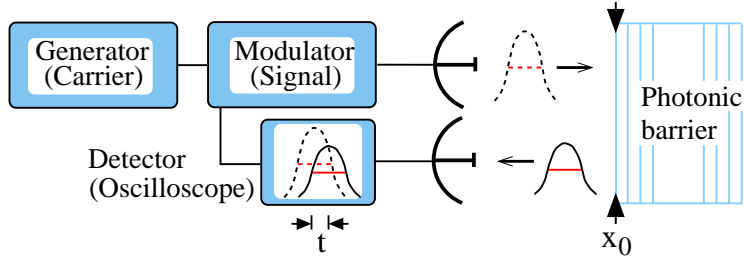}
 \caption{Experimental setup for reflection time measurement. 
A pulse--like signal of halfwidth $\Delta t = 8.5$~ns (corresponding to a 
bandwidth $\Delta f=80$~MHz) is modulated on a carrier in the microwave region $f_c=9.15$~GHz. 
The microwaves are transmitted and received by two parabolic antennas. The reflection times~$t$ for 
different photonic barriers are compared with the time of a reflection by a metallic mirror at the front 
surface of the barriers $x_0$, see Fig.~\ref{setup_structure}.\label{setup_signal}}
\end{figure}

\begin{figure}[hbt]
 \includegraphics[width=0.19\linewidth]{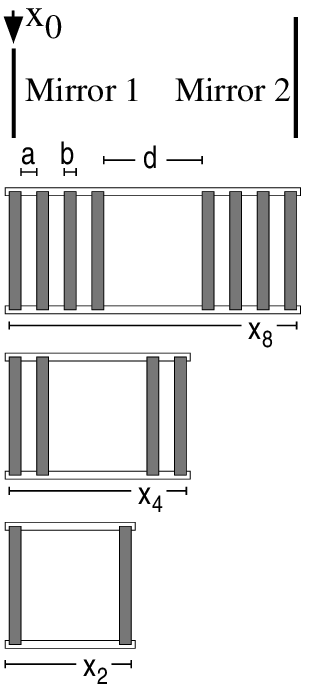}
\caption{Three photonic barriers of different total lengths $x_8 = 280$~mm, $x_4 = 226$~mm, and $x_2 = 199$~mm. 
Each structure consists of an alternating configuration of Perspex layers of width $b=5.0$~mm separated 
by air gaps $a=8.5$~mm. For certain frequencies the transmission of such a structure becomes exponentially 
damped by destructive interference so that the structure behaves like an opaque barrier, see Fig.~\ref{trans-sig}. 
The wide air gap $d=189$~mm allows to enlarge the barriers´ extention without increasing the attenuation.
Metallic mirrors at the front or back surface of the structure are used to simulate an ideal reflection.\label{setup_structure}} 
\end{figure}

The experimental setup and the investigated photonic barriers are sketched in 
Fig.~\ref{setup_signal} and~\ref{setup_structure}, respectively. For the time domain measurements, 
pulse--like signals with halfwidths of $\Delta t = 8.5$~ns, corresponding to a frequency--bandwidth 
of $\Delta f = 80$~MHz, were modulated 
on a high frequency carrier $f_c=9.15$~GHz produced by a microwave generator. Using the power output of the generator $P=1~$mW 
it can be estimated that each pulse is built by an ensemble of $P \, \Delta t / h\, f_c \approx 10^{12}$ single photons.
The microwave pulse was transmitted to the photonic barriers 
via a parabolic antenna; the reflected signal was received by a second antenna. A HP-54825 oscilloscope detected the 
envelope of the reflected microwave signal. The measurements were performed asymptotically, i.e. a coupling between generator, 
detector, and devices under test (photonic barriers or metallic mirrors) was avoided by the long optical distances of~3~m 
and by uniline devices in the microwave circuit. Due to the narrow radiation profile of the parabolic antennas of approximately~$5^\circ$ 
a direct coupling between them was excluded.

The barriers consist of two photonic lattices (periodic dielectric quarter wavelength structures) which are separated by 
an air gap, Fig~\ref{setup_structure}. Each lattice consists between one and four equidistant Perspex layers separated by air.
The refractive index of Perspex is $n=1.61$ in the measured frequency region.  
In order to build a photonic barrier for the microwave signal, the thicknesses of the Perspex $b=5.0$~mm and the air layers $a=8.5$~mm 
present a quarter of the microwave carrier´s wavelength in Perspex $\lambda_n = c/n f_c = 20.4$~mm and in air $\lambda_0 = c / f_c = 32.8$~mm, respectively.
At each surface of the Perspex layers a part $\rho =(n-1)/(n+1)$ of the incident wave or a factor $|\rho|^2 \approx 5\,\%$ 
of the incoming intensity is reflected. These reflections interfere constructively and result in a total  
reflection of nearly the same magnitude as the incident signal.
The air space $d=189.0$~mm between the two lattices
forms a cavity and extends the total length of the barrier. The resonance frequencies of the cavity
are given by multiples of $f_{\rm res} = c/2d = 794~$MHz. The frequency spectrum of the microwave signal lies completely 
in the nonresonant ´forbidden´ frequency region between the two resonances of the cavity at $11 \cdot f_{\rm res}$ and $12 \cdot  f_{\rm res}$ .

\begin{figure}[hbt]
\includegraphics[width=0.34\linewidth,angle=270]{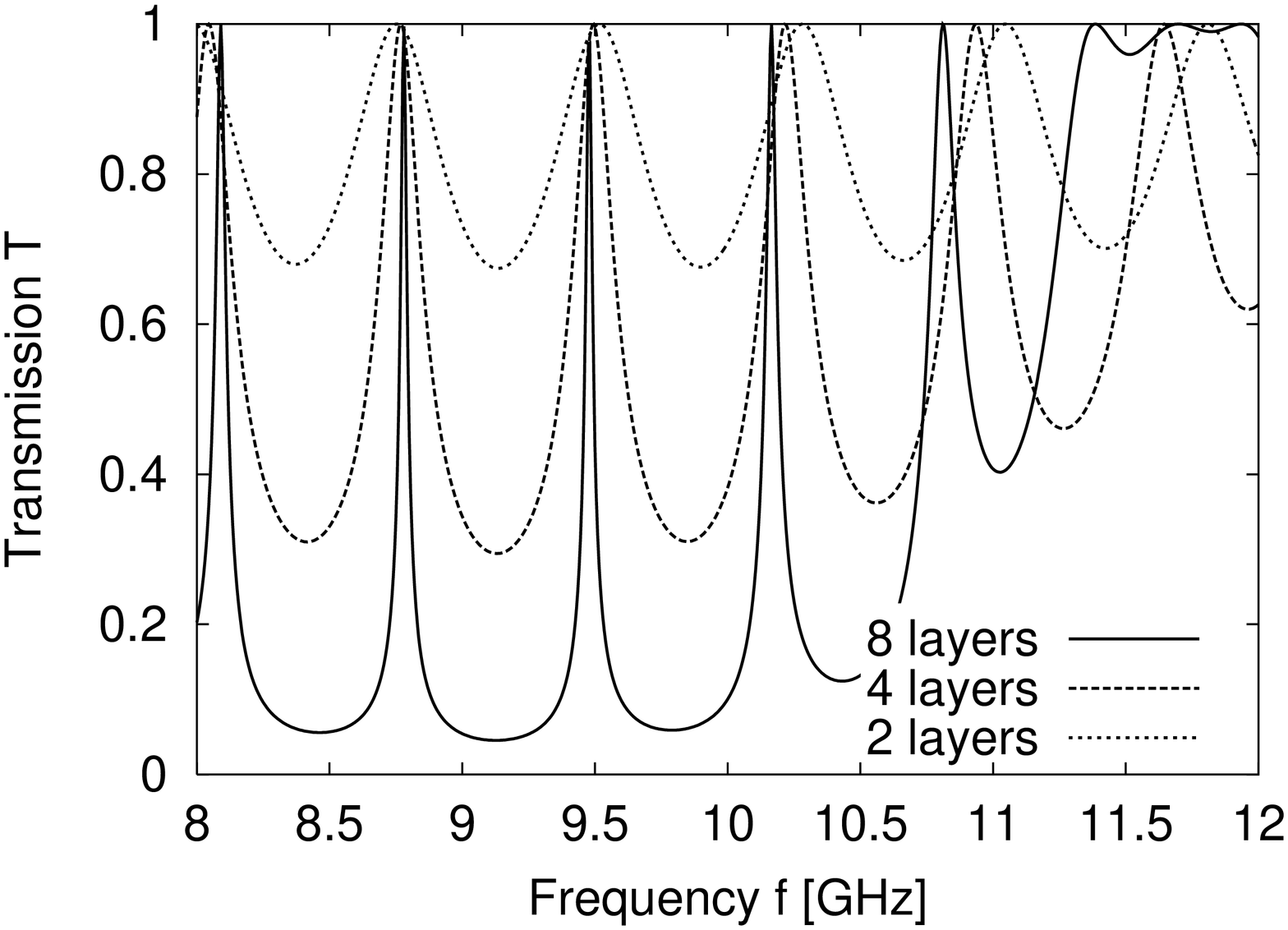}
\includegraphics[width=0.34\linewidth,angle=270]{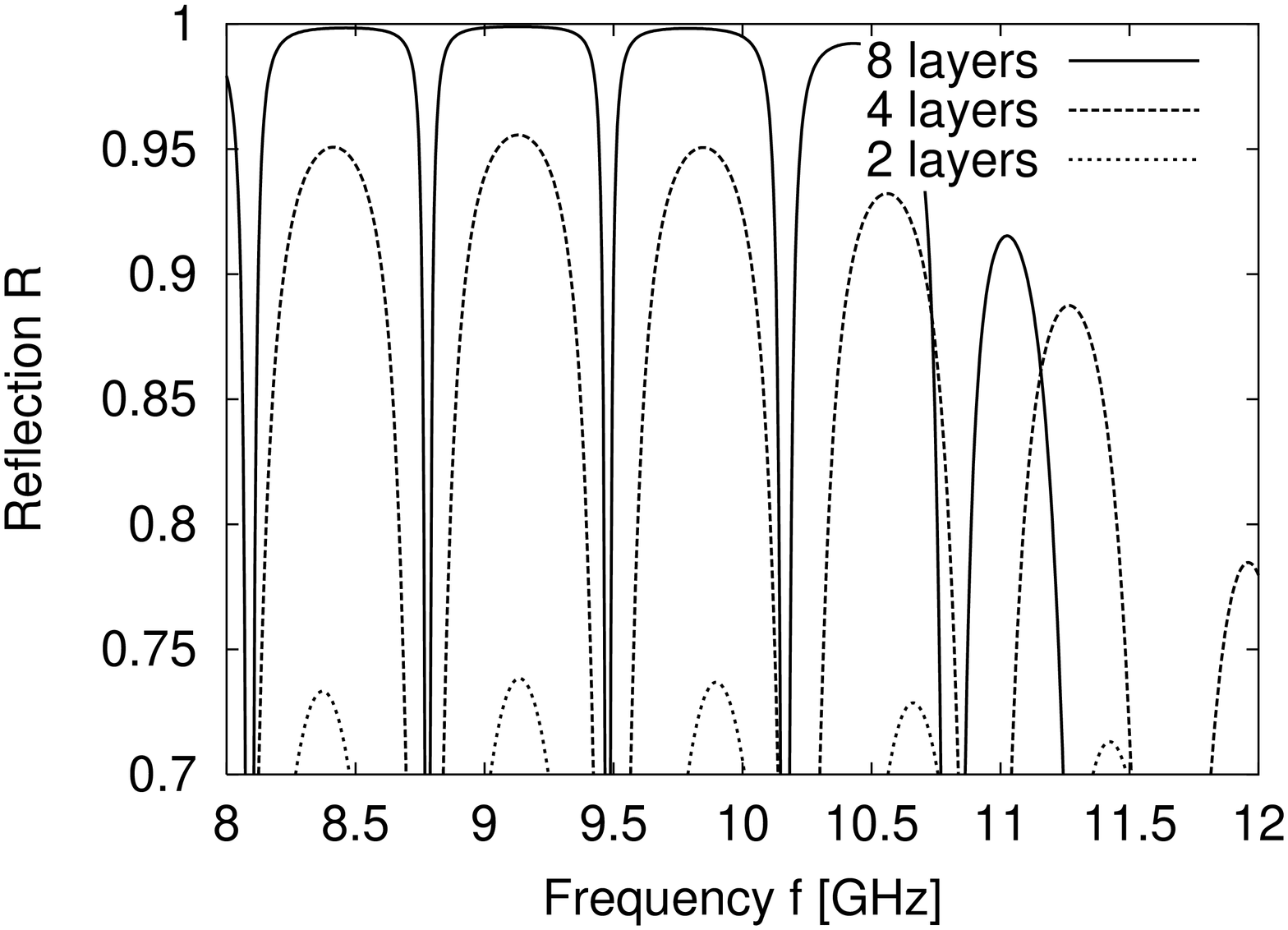}
\caption{Transmission $T$ (left) and reflection $R$ (right) for the photonic barriers consisting of 8, 4, and 2 layers of Perspex. 
         The frequency band $9.15~\mbox{GHz} \pm 40~\mbox{MHz}$ of the microwave pulse lay inside a 
         transmission gap where for the longest barrier only $T^2 = 0.25\,\%$ of the signal´s intensity is transmitted, while
         the rest of the pulse is reflected, according to the relationship $R^2 = 1 - T^2$.\label{trans-sig}}
\end{figure}

The calculated transmission and reflection of the barriers consisting of~8, 4, and 2 layers of Perspex are displayed in 
Fig.~\ref{trans-sig}. There are five pronounced forbidden bands separated by resonance transmission peaks 
of the cavity in the frequency range displayed. Within a frequency band of approximately
$9.15~\mbox{GHz}\pm 100~\mbox{MHz}$ around the carrier frequency~$f_c$ the complete 
structure behaves like a photonic barrier. Due to destructive interference the transmitted signal is exponentially 
attenuated by the number of Perspex layers.

\section{Partial reflection by photonic barriers}

For a normal dielectric medium with a real index of refraction $n > 1$, e.g. a sheet of glass, the propagation of 
the reflected microwave pulse is expected to be reshaped by partial reflections at the sheet's two surfaces. 
The maximum intensity of the reflection depends on the thickness of the sheet and varies sinusoidal~\cite{Feynman}. 
This behaviour is due to interference between waves reflected by the front and back surfaces of the single sheet.
We are now investigating the behaviour of the photonic barriers sketched in Fig.~\ref{setup_structure}, which have a purely imaginary index of refraction. 

\begin{figure}
\begin{center}
\includegraphics[width=0.5\linewidth,angle=270]{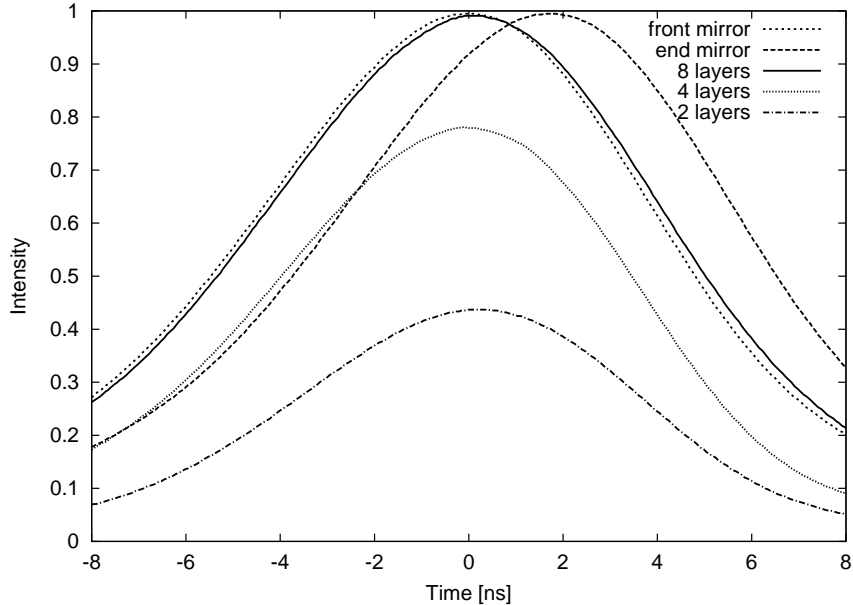}
\end{center}
\caption{Signals reflected by barriers of different lengths: An ideal reflection by a metallic mirror at the surfaces~$x_0$ of the 
                  barriers defines the time $t=0$, see Fig.~\ref{setup_structure}. An ideal reflection by a second mirror at the back surface $x_0 + x_8$ of the longest barrier 
                  is detected after the expected propagation time of approximately $2x_8/c \approx 1.9$~ns. The three other pulses were reflected by the barriers consisting 
                  of 8, 4, and 2~layers of Perspex. The time delay of the three reflected pulses keeps mainly constant
                  while the magnitudes of the signals depend on the number of Perspex layers. The short reflection 
                  time $t\approx$~100~ps corresponds to the tunnelling time $\tau\approx 1/f_c$ for a transmission through the barrier. 
                  A slightly larger delay time for the structure consisting of 2~layers indicates an insufficiently opaque barrier.
                  \label{gitter}}
\end{figure}

A signal sent to the metallic mirror, placed at the front surfaces~$x_0$ of the barriers, is reflected
and the reflected signal is detected by the oscilloscope after a certain time delay, Fig.~\ref{setup_signal}. We will 
subtract this time delay from all further measurements in order to use the arrival time of that pulse 
as a time reference $t=0$. Thus, a pulse reflected by a metallic mirror placed at the end of the barrier 
at $x_0 + x_8$ is expected to arrive at a time $t= 2\,x_8/c = 1.87$~ns, see Fig~\ref{gitter}.

The partial reflection by the photonic barriers revealed a strange
behaviour: if the length of the barrier was shortened from 8 to 4 or 2 layers (Fig.~\ref{setup_structure}), 
the time delay of the reflection kept constant whereas the amplitude decreased as a result of 
the increasing transmission (Fig.~\ref{trans-sig}). The measured time delay of the pulses reflected by the barriers differs 
approximately $t \approx 100$~ps from the reflection time at the front mirror $x_0$, see Fig.~\ref{gitter}. 
This delay time corresponds to the tunnelling time $\tau \approx 1/f_c$ for a signal in the microwave 
frequency range~$f_c=9.15$~GHz \cite{Haibel,esposito}.

To add further credibility to the time domain measurements, the reflection experiment was verified in
the frequency domain using guided microwaves and a network analyzer HP-8510. The photonic lattices were 
constructed from layers of Perspex inside X--band waveguides in an analogous arrangement to the above presented 
free space experiment~\cite{vetter}. The geometry of the structure ($a=12~$mm, $b=6~$mm, and $d=130~$mm) 
resulted in a forbidden band around $f_c=8.44~$GHz of width $\Delta f \approx 100$~MHz. 
Because the reflections at the Perspex layers inside the waveguide are stronger than in free space, 
the largest barrier consists of 6 layers of Perspex. To obtain a higher resolution we also used barriers with
odd numbers of 3 and 5~layers. As a result, also for these unsymmetrical barriers the transmission and refection time of a pulse 
did not depend on the side of incidence.

\begin{figure}
\begin{center}
\includegraphics[width=0.65
\linewidth,angle=0]{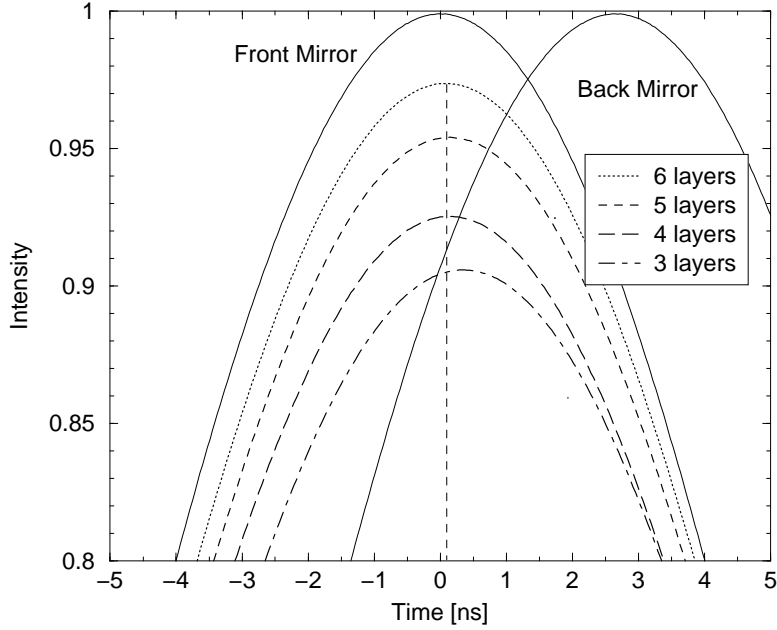}
\end{center}
\caption{Signals reflected by photonic barriers inside a waveguide consisting of 6 to 3 layers of Perspex. 
                  The solid pulses indicate the position of a reflection by metallic mirrors at the front and behind 
                  the largest photonic barrier of 6 layers with a total length of $x_6=214$~mm. The dashed pulses are the 
                  reflections at the barriers consisting of 6, 5, 4, and 3 layers of Perspex. The reflections of the barriers
                  were detected after a short time delay of approximately $t= 100~$ps, which equals the tunnelling 
                  time~$\tau$ (vertical line). The magnitude of the reflected pulses carried the information of
                  the length of the barrier in question ($x_5= 196$~mm, $x_4= 178$~mm, $x_3= 160$~mm).
                  \label{puls2}}
\end{figure}

After measuring the frequency spectra of the barriers for transmission and reflection, the propagation of pulses in the time domain
could be reconstructed by Fourier transforms. In order to simulate the reflection at a photonic barrier, the frequency components within the band gap at~$f_c$ was used to construct the pulses. 
Figure~\ref{puls2} shows the reconstructed pulses after a reflection by barriers of 3, 4, 5, and 6~layers. 
The frequency domain measurements confirm the above presented free space measurements: again the reflection time does not 
depend on the length of opaque barrier.

\section{Conclusions}

In both experiments the applied signal pulse had a carrier frequency~$f_c$ in the center of the
photonic barriers´ forbidden band gap and a narrow frequency--bandwidth~$\Delta f$ about~$1\,\%$ of~$f_c$.
Thus all frequency components of the signal were evanescent. In this case there is no finite 
phase--time or group delay expected nor observed for the wave packet inside a barrier~\cite{Hartman,nimtz2}. 
Such a behaviour seem to explain the experimental data of reflection by opaque barriers: Evanescent modes 
appear to be nonlocal at least up to some ten wavelengths as experiments have shown in this study.
The distance of observing nonlocality effects is limited by the exponential decay of the field 
intensity of evanescent modes, i.e. of the probability in the wave mechanical particle analogy. 

In measuring the reflection duration of wave packets by photonic barriers we observed that the partial reflection 
by the back surface has an instantaneous effect on the amplitude, whereas the reflection duration is 
not changed. Obviously the information on photonic barrier length is available at the barrier's front surface within the short 
tunnelling time. This is a strange property which Newton suggested erroneously to explain partial reflection of 
corpuscles by dielectric layers~\cite{Feynman}.

We thank H. Aichmann, P. Mittelstaedt, and A. Stahlhofen for helpful discussions, B.~Clegg for a critical reading 
of the manuscript, and M.~E.~Perel'man for giving us the paper on his investigation prior to publication. 

\end{document}